# Optical Spectra of Four Objects Identified with Variable Radio Sources

V. Chavushyan[1], R. Mujica[1], A. G. Gorshkov[2], V. K. Konnikova[2]\*, and M. G. Mingaliev[3]

[1] *National Institute of Astrophysics, Optics, and Electronics, Puebla, Mexico*
[2] *Sternberg Astronomical Institute, Universitetskiĭ pr. 13, Moscow, 119899 Russia*
[3] *Special Astrophysical Observatory, Russian Academy of Sciences, Nizhniĭ Arkhyz, Stavropol'skiĭ kraĭ, 357147 Russia*

**Abstract**—We obtained optical spectra of four objects identified with variable radio sources. Three objects (0029+0554, 0400+0550, 2245+0500) were found to be quasars with redshifts of 1.314, 0.761, and 1.091. One object (2349+0534) has a continuum spectrum characteristic of BL Lac objects. We analyze spectra of the radio sources in the range 0.97–21.7 GHz for the epoch 1997 and in the range 3.9–11.1 GHz for the epoch 1990, as well as the pattern of variability of their flux densities on time scales of 1.5 and 7 years. © *2000 MAIK "Nauka/Interperiodica".*

## INTRODUCTION

We obtained optical spectra of four objects (0029+0554, 0400+0550, 2245+0500, 2349+0534) from a complete flux-limited sample of Zelenchuk-survey radio sources [1]. The sample contains all sources with fluxes $S > 200$ mJy at a frequency of 3.9 GHz and covers 24 h in right ascension at declinations 4°–6° and $|b| > 10°$ [2]. Since the limiting flux of the sample is low enough, the entire luminosity function of quasars is observed up to redshifts $z \approx 1$, which allows an attempt to be made to detect cosmological evolution of the quasar luminosity function after determining the redshifts for all sample objects.

## OPTICAL OBSERVATIONS

We carried out optical observations in October 1998 with the 2.1-m telescope at the Guillermo Haro Observatory in Cananea of the National Institute of Astrophysics, Optics, and Electronics, Mexico (INAOE). We used the LFOSC spectrophotometer equipped with a 600 × 400-pixel CCD array [3]. The detector readout noise was 8 e⁻, and the wavelength range covered was 4200–9000 Å with a 8.2-Å dispersion. The effective instrumental resolution was ~16 Å.

We performed the standard reduction procedure—the removal of cosmic-ray hits, bias and flat-field corrections, wavelength linearization, and flux calibration—by using the IRAF package.

The source 0440+0550 was observed with a 60-min exposure; the exposure time for the remaining objects was 40 min. Magnitudes were taken from the Automated Plate Scanner Catalog of the Palomar Sky Survey [4].

## RADIO OBSERVATIONS

We observed all four radio sources yearly from 1984 until 1992 at frequencies of 3.9 and 7.5 (or 7.7) GHz with the RATAN-600 Southern Sector. In 1990, spectra of the sources at frequencies of 3.9, 4.8, 7.7, and 11.1 GHz [5] were obtained with the Southern and Northern Sectors. Since 1996, the sources have been observed two or three times a year simultaneously at six frequencies (0.97, 2.3, 3.9, 7.7, 11.1, and 21.7 GHz) with the Northern and Western RATAN-600 Sectors. Detector parameters and beam characteristics for the Northern and Western RATAN-600 Sectors are presented in [6, 7]. The same characteristics for the Southern Sector are given in [5]. In every series, the sources were observed 10 to 15 times each. The source flux was obtained by averaging all data in each series. Flux errors were determined from the scatter of fluxes detected daily in a given series; they include all types of error: noise, calibration error, calibration-signal referencing error, antenna pointing error, etc. The reduction procedure is described in [8]. The flux-density scales in different years were reduced to the scale adopted in [7], which presents the observations of all sample sources with power-law spectra.

## RESULTS

Table 1 gives objects' names, their radio positions, and differences between the optical and radio positions. The first column contains the source names consisting of hours and minutes of right ascension and degrees and minutes of declination for the epoch 2000. The source names for the epoch 1950 appearing in previous papers are given in parentheses.

---

\* E-mail address for contacts: algor@sai.msu.su

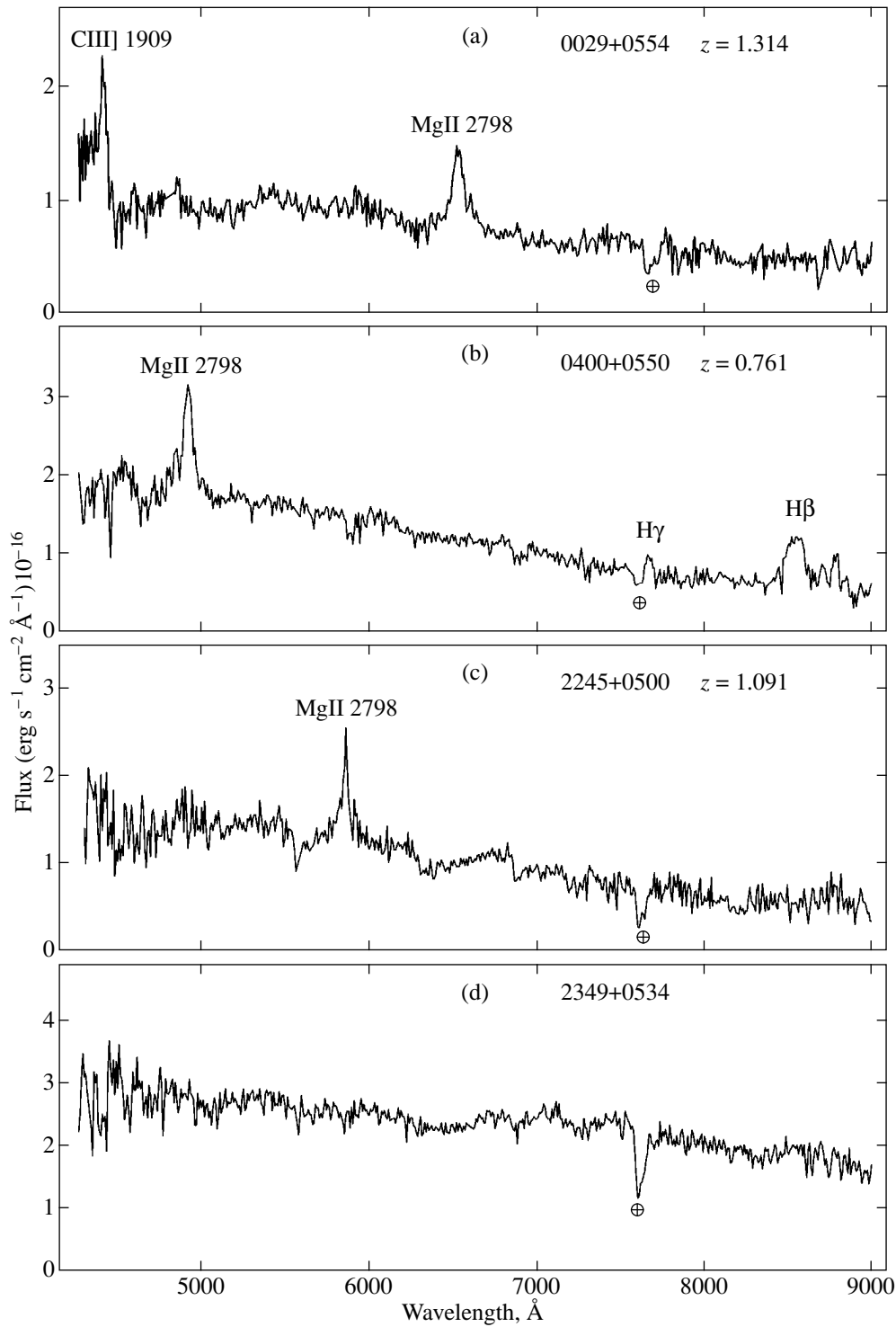

**Fig. 1.** Optical spectra of the objects identified with the radio sources 0029+0554, 0400+0550, 2245+0500, and 2349+0534.

The radio positions were taken from the JVAS2 catalog of 2118 northern-sky compact radio sources [9]; the rms error of the positions in this catalogue is 0.014 arcsec. The error in the positions of the optical objects is 0.5 arcsec [4].

The difference between the radio and optical positions for all sources is smaller than $3\sigma$ of the total error of the radio and optical positions.

**The source 0029+0554** was identified with a starlike object [10]. Figure 1a shows the object's optical spectrum. The spectrum exhibits two intense lines, which can be interpreted as the semi-forbidden C III] 1909 Å line and the Mg II 2798 Å line at redshift $z = 1.314$. The spectrum is typical of quasars. The RATAN-600 observations of the source show that its flux density slowly increases: between 1990 and 1997, the flux density increased from 300 to 480 mJy at 3.9 GHz and from 360 to 498 mJy at 7.5 GHz. Figure 2a presents the spectra of 0029+0554 obtained in August 1990 in the frequency range 3.9–11.1 GHz and in August 1997 in the range 0.97–21.7 GHz. Both spectra are rising, with the two-frequency spectral index between 3.9 and 7.5 GHz in both spectra being $\alpha \approx 0.23$ ($S \propto \nu^\alpha$). In the August 1997 spectrum, the flux density reaches a maximum at 22.7 GHz. During 1.5 years, the source was observed at six frequencies in four series and exhibited no statistically significant variability of its flux density. The relative variability amplitude $V$ was calculated by the method described in [11]. On a time scale of ~8 years (1984–1992), the relative amplitude of long-term variability is 0.11 and 0.14 at 3.9 and 7.5 GHz, respectively [2]. All radio observations show that the flux density of the radio source 0029+0554 varies rather slowly, and the development of a single outburst at the stage of electron acceleration and/or magnetic-field strengthening is observed. The absolute spectral radio luminosity of the source 0027+056 at the frequency of maximum in a homogeneous isotropic cosmological model with a zero cosmological constant, deceleration parameter $q_0 = 0.5$, and $H = 50$ km s$^{-1}$ Mpc$^{-1}$ is $L_\nu = 2.5 \times 10^{34}$ erg s$^{-1}$ Hz$^{-1}$.

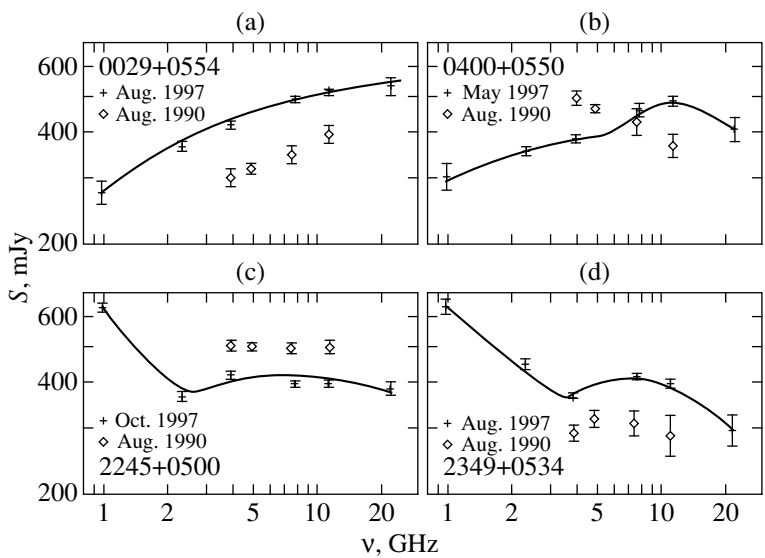

**Fig. 2.** Radio spectra of the sources 0029+0554, 0400+0550, 2245+0500, and 2349+0534 in 1990 and 1997.

**The source 0400+0550** was identified with a starlike object [12]. The source's optical spectrum (Fig. 1b) exhibits three lines, which are interpreted as Mg II 2798 Å, H$\gamma$ 4340 Å, and H$\beta$ 4861 Å at redshift $z = 0.761$. The spectrum is typical of quasars. Figure 2b presents the spectra of the source 0400+0550 obtained in August 1990 in the range 3.9–11.1 GHz and in May 1997 in the range 0.97–21.7 GHz.

The August 1990 spectrum is falling with the mean power-law index $\alpha = -0.23$, typical of the outburst spectrum in an optically thin spectral region. The May 1997 spectrum is composite; it can be separated into two components: an extended one with a power-law spectrum and a compact one with a peak in the spectrum attributable to synchroton self-absorption at 13.63 GHz, i.e., the initial stage of development of a new outburst is observed. During the source's observations from April 1996 until May 1997, no statistically significant variability of the flux density was found.

**Table 1.** Source names, radio positions, and differences between radio and optical positions

| Source name | Radio positions, J2000.0 | | Radio–optical | |
|---|---|---|---|---|
| | R.A. | Decl. | R.A., arcsec | Decl., arcsec |
| 0029+0554 (0027+056) | 00$^h$29$^m$45$.^s$90 | +05°54′40″69 | −0.30 | 0.09 |
| 0400+0550 (0357+057) | 04 00 11.74 | +05 50 43.14 | 0.0 | −0.46 |
| 2245+0500 (2243+047) | 22 45 53 65 | +05 00 56.96 | −1.35 | −1.04 |
| 2349+0534 (2346+052) | 23 49 21.06 | +05 34 39.85 | 0.15 | −0.55 |

**Table 2.** A summary of the results of optical and radio observations

| Object | Lines | Å | Type | $z$ | $B$ | $R$ | $S$, mJy $\frac{3.9, 11.1}{1997}$ | $V$ $\frac{3.9, 7.5}{1984-1991}$ |
|---|---|---|---|---|---|---|---|---|
| 0029+0554 | $\frac{\text{C III]}}{\text{Mg II}}$ | $\frac{1909}{2798}$ | QSO | 1.314 | 18.7 | 18.3 | $\frac{422\ (9)}{524\ (6)}$ | $\frac{0.11}{0.14}$ |
| 0400+0550 | $\frac{\text{Mg II}}{\frac{\text{H}\gamma}{\text{H}\beta}}$ | $\frac{2798}{\frac{4340}{4861}}$ | QSO | 0.761 | 18.1 | 17.5 | $\frac{387\ (9)}{497\ (13)}$ | $\frac{0.12}{0.13}$ |
| 2245+0500 | Mg II | 2798 | QSO | 1.091 | 18.6 | 18.4 | $\frac{405\ (8)}{399\ (8)}$ | $\frac{0.34}{0.41}$ |
| 2349+0534 | none | | BL Lac | | 19.0 | 18.0 | $\frac{366\ (11)}{400\ (10)}$ | $\frac{0}{0}$ |

The relative amplitude of long-term variability in seven years is $V = 0.13$ and $0.15$ at 3.9 and 7.5 GHz, respectively [2]. By contrast to the previous source, 0400+0550 exhibits variations on a time scale of 2 or 3 years. The source's absolute spectral radio luminosity at the frequency of maximum of the May 1997 spectrum is $L_\nu = 8.6 \times 10^{33}$ erg s$^{-1}$ Hz$^{-1}$.

**The source 2245+0500** was identified with a starlike object [13]. It was observed in September 1989 with the 1-m telescope of an expedition of the Institute of Theoretical Physics and Astrophysics (Lithuanian Academy of Sciences) at Mount Maidanak (Uzbekistan) in $U$, $B$, and $V$. Its $U$ magnitude was $19.3 \pm 0.5$, $U-B = 0$, and $B-V = 0.3$ [13].

Our optical spectrum exhibits a single line (Fig. 1c). Judging by its profile and intensity, this is the Mg II 2798 Å line at redshift $z = 1.091$, i.e., the object belongs to quasars. Its $B$ and $R$ magnitudes are 18.6 and 18.4, respectively. The source may also have a variable optical flux.

Figure 2c shows the source's snap-shot spectra obtained in August 1990 and in November 1997. Both spectra have the spectral index $\alpha = 0$ virtually in the entire centimeter range; however, subtracting the extended component from the 1997 composite spectrum reveals a compact component with a peak flux density at 15 GHz. The source was observed at six frequencies from July 1996 through February 1998; the relative variability amplitude $V$ at 3.9, 7.7. and 11.1 GHz was, respectively, 0.11, 0.32, and 0.38. The relative amplitude of long-term variability in [2] was 0.34 and 0.41 at 3.9 and 7.5 GHz, respectively.

The daily RATAN-600 observations in January–February 1998 revealed no rapid variability of the source on time scales shorter than 30 days.

The absolute spectral luminosity of the source at the frequency of maximum in the compact component's spectrum is $L_\nu = 1.25 \times 10^{34}$ erg s$^{-1}$ Hz$^{-1}$.

**The source 2349+0534** was identified with a starlike object [13]. The object's optical spectrum (Fig. 1d) is a purely continuum one without noticeable lines. The spectrum is characteristic of BL Lac objects. The optical spectrum was apparently obtained during a radio outburst of the source. During the observations from 1984 until 1992, the source's flux density was essentially the same in all observations, the mean flux densities at 3.9 and 7.5 GHz were, respectively, 300 and 305 mJy, and the spectral index was virtually zero in all observations.

After a considerable break, the source was observed at six frequencies from April 1996 through August 1997. The flux density increased: we apparently observed the initial stage of an outburst in early 1996. Figure 2d shows the source's spectra for the epochs August 1990 and August 1997.

The August 1997 spectrum is composite and can be separated into two components: an intense extended one with a power-law spectrum ($\alpha = -0.9$) and a compact one with a peak in the spectrum at 10.7 Ghz.

## CONCLUSION

The results of our optical observations and some characteristics of the objects in the radio band are presented in Table 2.

Of the four objects whose optical spectra we obtained, three are quasars and one is a BL Lac object. For all sources, the ratio of radio and optical flux densities lies in the range $(2-4) \times 10^3$.

## ACKNOWLEDGMENTS

We wish to thank J. Miramon, G. Miramon, R. Gonzales, and S. Noriega for technical support and help in the observations with the 2.1-m telescope. This study was supported by the Russian Foundation for Basic Research (project no. 98-02-16428), the Program "Universities of Russia" (project no. 5561), the State Science and Technology Program "Astronomy" (project no. 1.2.5.1), and in part by a CONACYT grant (no. 28499-E).

*Translated by N. Samus'*